\documentclass[twocolumn,preprintnumbers,amsmath,amssymb,floatfix,showpacs,prl]{revtex4}

\usepackage{dcolumn}
\usepackage{bm}

\begin{document}


\title{
Berry Curvature on the Fermi Surface: Anomalous Hall 
Effect as a Topological Fermi-Liquid Property
}

\author{F. D. M. Haldane}
\affiliation{Department of Physics, Princeton University,
Princeton NJ 08544-0708}

\date{Received 28 June 2004; revised manuscript received  20 October 2004}

\begin{abstract}
The intrinsic anomalous Hall effect in metallic ferromagnets is 
shown to be controlled by Berry phases accumulated by adiabatic
motion of  quasiparticles on the Fermi surface, 
and is purely a Fermi-liquid property, not a
``bulk'' Fermi sea property like  Landau diamagnetism,
as has been previously supposed.
Berry phases are a new
topological ingredient that must be added to Landau Fermi-liquid
theory in the presence of
broken inversion or time-reversal symmetry.
\end{abstract}


\maketitle
Renewed interest in the anomalous Hall
effect (AHE) 
in metallic ferromagnets has lead to a reinterpretation
of the classic
Karplus-Luttinger formula\cite{LK} for the ``anomalous velocity''
in terms
of the Berry curvature of occupied electronic 
Bloch states\cite{jung,nag}. 
This gives an {intrinsic} contribution
to the  Hall conductivity in the
low-temperature
clean limit of metallic ferromagnets
when the quasiparticle lifetimes become long, and now appears
to be the dominant contribution to the AHE\cite{fang,ong}.

The expression\cite{LK,jung,nag} for
the intrinsic Hall conductivity of metals with broken time-reversal
symmetry
has the all the appearance of a ``bulk'' bandstructure
property that depends on {all} the filled electronic states, not just
the ones at the Fermi level.    However, this  seems 
at odds with the spirit of Landau's Fermi-liquid theory, which holds
that charge transport in metals involves only quasiparticles
with energies within $k_BT$ of the Fermi level. 
      
In this Letter, I show that,
despite appearances,  the \textit{nonquantized part}
of the intrinsic Hall conductivity 
is {completely expressible in terms of
Berry phases}\cite{berry} 
{of quasiparticles moving on the Fermi surface}, and thus
fully consistent with Fermi liquid theory.  This exposes
a new topological ingredient that must be added to
Fermi-liquid theory unless both inversion and time-reversal symmetry
are present: {quasiparticle Berry phases}.

The ``anomalous velocity''
is an extra term\cite{niu, marder} in 
the semiclassical equations of motion of a Bloch electron
in weak electric and magnetic fields; ignoring Zeeman couplings, these  are
\begin{eqnarray}
\hbar \frac {dk_a}{dt} &=& eE_a(\bm x) + eF_{ab}(\bm x)\frac{dx^b}{dt}, \\
\frac {dx^a}{dt} &=& \frac{1}{\hbar}\nabla^a_k\varepsilon_n(\bm k)
+ \mathcal F_n^{ab}(\bm k)\frac{dk_b}{dt}, 
\label{aha}
 \end{eqnarray}
where $F_{ab}$ = $\nabla_a A_b - \nabla_b A_a$ 
= $\epsilon_{abc}B^c$
is the magnetic flux density written as an antisymmetric  tensor,
$\nabla_a$ $\equiv$ $\partial/\partial x^a$,
$\varepsilon_n(\bm k)$ is the energy of a Bloch electron in band $n$,
$\nabla_k^a$ $\equiv$ $\partial/\partial k_a$
and $\mathcal F_n^{ab}$ is the antisymmetric ``Berry curvature'' tensor
in $\bm k$-space, described below. The anomalous
velocity in (\ref{aha}) is the $\bm k$-space dual of the
Lorentz force.

If one writes the electron occupations  
$\langle n_{k n} \rangle $ as
$ n^0_{n}(\bm k,\mu)$ + $\delta n_{kn}$, 
where $n^0_{ n}(\bm k,\mu)$ is the ground state occupation
function  at chemical
potential $\mu$,
the linear
current response to
 a uniform electric field $\bm E$ (with $\bm B$ = 0)
is
\begin{equation}
J_e^a = \frac{e}{\hbar}\frac{1}{\Omega N}\sum_{k n} 
\nabla_k^a \varepsilon_n(\bm k) \delta n_{kn}
+ \sigma_0^{ab}(\mu)E_b,
\end{equation}
where $N$ is the number of primitive unit cells, which
have volume $\Omega$.
Here $\sigma_0^{ab}(\mu)$ is an intrinsic ground-state property
describing a dissipationless Hall conductivity:
\begin{equation}
\sigma_0^{ab}(\mu) = 
 \frac{e^2}{\hbar}\frac{1}{\Omega N}\sum_{ k n}\mathcal F^{ab}_n n^0_{n}(\bm k,\mu)
\label{qhe}
\end{equation}
This also controls
the low-temperature limit of the thermal Hall conductivity
$\kappa^{ab}$ (Righi-Leduc effect) and the Peltier coefficient
$\alpha^{ab}$; their ``intrinsic'' parts are
\begin{equation}
\kappa^{ab}_0(\mu) = \frac{\pi^2}{3}\frac{k_B^2T}{e^2}\sigma_0^{ab}(\mu),
\\
\quad
\alpha^{ab}_0(\mu) = e \frac{\partial \kappa_0^{ab}(\mu)}{\partial \mu}
 .
\end{equation}
A heat current $J^a_Q$ = $T\alpha_0^{ab}E_b$
flows with  the 
Hall current.

 If time-reversal symmetry is present, the electronic bands have the property
\begin{equation}
\varepsilon_n(-\bm k)  = \varepsilon_n(\bm k),
\quad
\mathcal F^{ab}_n(-\bm k) = - \mathcal F^{ab}_n(-\bm k),
\end{equation}
the sum (\ref{qhe}) cancels,
and the intrinsic Hall conductivity vanishes.   If
inversion symmetry is also unbroken,
$\mathcal F^{ab}_n(-\bm k)$ = $\mathcal F^{ab}_n(\bm k)$, and
the Berry curvature vanishes.

The Berry curvature is obtained from a ``vector potential''
derived from the 1-particle Bloch states $|\psi_n(\bm k)\rangle$:
\begin{eqnarray}
&&\mathcal A_n^a(\bm k) = -i \langle \psi_n(\bm k)|\nabla^a_k\psi_n(\bm k)\rangle,
\\
&&\mathcal F^{ab}_n(\bm k)
= \nabla_k^a\mathcal A^b_n(\bm k) - \nabla_k^b\mathcal A^a_n (\bm k) ,
\label{vp} \\
&&\epsilon_{abc}\nabla_k^a\mathcal F_n^{bc}(\bm k) 
= \sum_i q_{ni} \delta^3(\bm k-\bm k_{ni}), \quad q_{ni} = \pm 2\pi.\qquad
\end{eqnarray}
The last equation is the divergence of the $\bm k$-space
Berry  curvature field 
$\mathcal F^{ab}_n(\bm k)$: it is divergence-free except for quantized
``monopole'' sources with a ``charge quantum'' $2\pi$, which are associated
with band degeneracies.  These occur at isolated $\bm k$ points:
in complex Hermitian
eigenproblems, it is sufficient to vary three parameters (here, 
the components of $\bm k$) to encounter degeneracies.

Assuming that the one-electron
energy $\varepsilon_n(\bm k)$ is nondegenerate, the wavefunctions
are completely defined, {
except for an arbitrary phase factor}.
If both time-reversal
and inversion symmetry are present, this can consistently be chosen real,
but otherwise, is an arbitrary complex factor
that can vary continuously with $\bm k$. 
The Berry vector potential $\mathcal A^a_n(\bm k)$
depends on this ``gauge choice''
but the curvature
$\mathcal F^{ab}_n(\bm k)$ is a well-defined
gauge-invariant quantity with physical significance.
The Berry phase\cite{berry} for a {closed} path $\Gamma$ is also
gauge invariant:
\begin{equation}
\exp i \phi_n(\Gamma ) = \exp i \oint_{\Gamma} \mathcal 
A^a_n(\bm k)
 dk_a .
\end{equation}
The integral is the curvature flux linked through $\Gamma$:
the $2\pi$ ambiguity of the Berry phase
quantizes the ``charge''
of the monopole sources  of the curvature field.

In a 3D bandstructure, the integral over
the Brillouin zone (BZ) of the Berry curvature of
a nondegenerate band is a topological invariant\cite{halp}
that is a generalization of the better-known 2D 
{Chern number}\cite{TKNN,Avron}
\begin{equation}
\frac{1}{2\pi}
\int d^3\bm k
\mathcal F^{ab}_n(\bm k) P_{\text{BZ}}(\bm k)= C_n
\epsilon^{abc}G^C_{cn},
\end{equation}
where $C_n$ is an integer Chern number, $1/2\pi$ times the integral
of the Berry curvature over a compact 2D surface, {e.g.},
a 2D BZ (a torus), 
and $\bm G^C_n$
is a primitive reciprocal lattice vector;
$P_{\text {BZ}}(\bm k)$ = 1 inside the BZ, 0 outside it. 

The intrinsic Hall conductivity may be parameterized as
$\sigma^{ab}_0$ = $(e^2/\hbar)\epsilon^{abc}(K_c/(2\pi)^2)$, where
$\bm K$ is dimensionally a wavevector:  in one-electron band
theory,
\begin{equation}
\frac{1}{2\pi}
\sum_n \int d^3 \bm k\, \mathcal F^{ab}_n 
P_{\text{BZ}}(\bm k)n^0_n(\bm k,\mu)
= \epsilon^{abc} K_{c}.
\label{3dqhe}
\end{equation} 
If band $n$ is completely below the Fermi level, it contributes
a quantized amount $C_n\bm G^C_n$ to $\bm K$.
This produces an integer quantum Hall
effect (QHE) with ``filling factor'' $\nu$ = $C_n$ on the 
lattice planes indexed by $\bm G^C_n$.

The QHE is usually discussed in the context
of  strong magnetic fields where the
electronic states are split up into Landau levels,
so it might be wondered how ``simple'' Bloch electrons
could exhibit a QHE without Landau levels.  In fact, only broken
time-reversal symmetry is required: the possibility of a $\bm B$ = 0 
``zero-field QHE'' was first demonstrated in Ref.\cite{haldane88},
using a model that, in retrospect, exhibits both a non-metallic
QHE phase {and} a metallic AHE phase.

It will be useful to understand the process by which the Chern 
invariants 
of a 3D bandstructure can change. 
While a band remains non-degenerate, its Chern invariant
is ``quantized'' to be a  reciprocal lattice vector.   As a control
parameter is varied, two bands may come into contact at some point
in the BZ,
and this initial degeneracy point then subsequently splits into
two ``Dirac point'' singularities
(near which the energy dispersion is linear).  The bands are now tightly
coupled by a ``Berry flux loop'' where Berry curvature flux $2\pi$ passes 
from one band to the other through  one Dirac point, then  returns
through the other one.   There is a striking analogy to the idea
of ``wormholes'' connecting different universes, where here the two
``universes'' are Bloch bands, and ``space''' is $\bm k$ space.
Each band has one positive and one negative
monopole source of Berry curvature; at each Dirac point
the two bands have opposite-sign sources.
Eventually, after relative displacement by a 
reciprocal lattice vector $\bm G$, the monopoles may recombine, 
allowing the bands to split apart.
This process conserves the sum of their invariants, but individually
they change by $\pm \bm G$.

It is useful to first examine (\ref{qhe}) in the
simpler 2D case. Then $\sigma_0^{xy}$ = $\nu e^2/(2\pi \hbar)$,
where $\nu$ = $\sum_n \nu_n$, and
\begin{equation}
\nu_n(\mu) = \frac{1}{2\pi}\int d^2{\bm k} \,\mathcal F^{xy}_n(\bm k)
P_{\text{BZ}}(\bm k)n_n(\bm k,\mu).
\end{equation}
For simplicity, assume that the occupied region does
not touch the Brillouin zone boundary (BZB), and drop $P_{\text{BZ}}(\bm k)$.
Using the Berry vector potential representation
and  integrating by parts gives
\begin{equation}
\nu_n = \frac{1}{2\pi} \int d^2{\bm k} 
\, \left (\mathcal  A^x_n\nabla_k^yn_n^0(\bm k)
- A^y_n\nabla_k^xn_n^0(\bm k)\right ) ,
\end{equation}
which is clearly a Fermi surface integral if the band is partially filled, 
since $n^0_n(\bm k)$ has a step discontinuity at the Fermi surface, and
is constant everywhere else.    If the Fermi surface is a simple
closed loop, this can be recognized as the integral giving the
Berry phase $\phi_F$ for an adiabatic path around the Fermi surface
(the $\bm k$-space version of the  Bohm-Aharonov effect)
\begin{eqnarray}
&&\nu_n = \frac{1}{2\pi}\oint  \mathcal A^a_n(\bm k_F) d \bm k_{Fa} 
 = \frac{\phi_F}{2\pi},
\end{eqnarray}
Since the Berry phase is ambiguous
by a multiple of $2\pi$, only the {nonquantized} part of
the intrinsic Hall conductivity is determined at the Fermi surface.
If the system has  evolved adiabatically 
along some path in a
parameter space
from one with
time-reversal symmetry, the QHE can be determined from
the history of $\phi_F$ during that process.

Armed with the insight that the AHE is a Fermi surface property,
I now examine the 3D problem.  In general, there may be multiple
sheets $S_{\alpha}$ 
of the Fermi surface, with both simple and multiply connected
topology, with pieces ``glued together'' at degeneracy points where
a line of high symmetry intersects the surface, or along lines where
a plane of high-symmetry intersects it; if time-reversal symmetry
is unbroken, there may also be Kramers degeneracy\cite{kramers}.
Complexes of intrinsically connected
sheets $\bm k^{(\alpha)}_F(\bm s)$, where $\bm s$ = $\{s^1,s^2\}$ 
is a surface parameterization,
will be referenced by a single  label $\alpha$, specified
implicitly if $\bm s \in S_{\alpha}$.
The outward normal unit vector $\hat{\bm n}(\bm s)$
is also the direction of 
the Fermi velocity.

The key Fermi surface property is the Luttinger sum rule
relating particle number to Fermi surface  volume.   
The Fermi surface geometry fixes the particle density modulo
integer multiples of the 
``density quantum'' $\rho_0$ = $1/\Omega$ associated with filled
bands.  The change in
particle density $\delta \rho_{\alpha}(\bm r)$ associated
with a local fluctuation $\delta \bm k_F(\bm r,\bm s)$ 
= $\bm \nabla \varphi (\bm r,\bm s)$ (where
$\varphi (\bm r,\bm s)$ is the quasiparticle
phase)
is just
proportional to the $\bm k$-space volume swept out by
the changing  Fermi surface:
\begin{eqnarray} &&
\partial_{\mu}\bm k_{F}(\bm s) \times
\partial_{\nu}\bm k_{F}(\bm s) =
 \ell_{\mu\nu}(\bm s) \hat {\bm n}(\bm s) ,
\\
&&\delta \rho_{\alpha}(\bm r) = 
\int_{S_{\alpha}} \frac{ds^{\mu}\wedge ds^{\nu}}{(2\pi)^3} \,
\ell_{\mu\nu}(\bm s)
\hat {\bm n}(\bm s)\cdot
\bm \nabla \varphi(\bm r,\bm s),\qquad \\
&&
\int_{S_{\alpha}} \frac{ds^{\mu}\wedge ds^{\nu}}{(2\pi)^3} 
\ell_{\mu\nu}(\bm s) 
\hat {\bm n}(\bm s)
= 
\rho_0 \sigma_{\alpha}
\bm R_{\chi}.
\end{eqnarray}
Here $\partial_{\mu}$ $\equiv$ $\partial/\partial s^{\mu}$;
a nonzero integer  $\sigma_{\alpha}$
signals a {``chiral anomaly''}: the system then has a
quasi-1D character where the 
primitive real-space lattice vector
$\bm R_{\chi}$ 
defines a special direction of lattice lines along which 
quasi-1D electrons predominantly move (such a band structure
may also have nonchiral Fermi-surface sheets.)

The change in density
$\delta \rho_{\alpha}$
if the Fermi surface sheet $\alpha$ is rigidly displaced by a constant
shift $\delta \bm k_F$ is $\sigma_{\alpha}
\rho_0 \bm R_{\chi}\cdot \delta \bm k_F$.
Note that the absolute value of
$\bm k_F(\bm s)$ is not invariant under position-space
gauge transformations $\bm k_F $ $\rightarrow$
$\bm k_F - (e/\hbar)\bm A(\bm r)$
 and only the {relative} displacements
(modulo reciprocal lattice vectors)
between Fermi vectors are physically meaningful.  
Gauge invariance requires that the total Fermi surface
chiral anomaly must vanish.
Typical Fermi surface sheets are nonchiral
but the possibility of chiral sheets
needs to be kept in mind; the integer
$\sigma_{\alpha}$ is a measure of how many distinct
chiral sheets are glued together.

It is straightforward to repeat the
integration by parts to expose the nonquantized part of the
3D  intrinsic Hall conductivity as a Fermi surface
property. 
The  Berry vector potential and 
associated gauge-invariant Berry curvature
for paths restricted to lie {in} a surface $\bm k(\bm s)$
are
\begin{equation}
\mathcal A_{\mu}(\bm s) =
\mathcal  A^{a}(\bm k({\bm s})) \partial_{\mu}
k_a(\bm s), \quad\mathcal F_{\mu\nu} = 
\partial_{\mu}\mathcal A_{\nu}
- \partial_{\nu}\mathcal A_{\mu} .
\end{equation}

The surface integral
that results from integration by parts of the
band-$n$ contribution is over a 
surface $S_n$ that is divided into a set of
one or more  outward-oriented
compact surfaces 
enclosing  occupied regions within the BZ:
\begin{eqnarray}
\bm  K_n
 &=& \frac{1}{2\pi}
\int_{S_{n}}ds^{\mu}\wedge ds^{\nu} 
\mathcal F_{\mu\nu}
(\bm s)\bm k(\bm s). 
\label{central}
\end{eqnarray}
Some parts of $S_n$ are on the Fermi surface, but there may
also be contributions from states below the Fermi level
which are on the BZ boundary (BZB).
The Fermi surface of band $n$ will be
divided into one or more disjoint oriented
surfaces
$S_{n\alpha}$, and the intersections of these with the BZB define
a set of conjugate pairs of closed directed paths $C_{n\alpha i\pm}$
which have Berry phases $
\pm\phi_{n\alpha i}$; 
the
``$+$'' path is displaced relative
to its partner on the opposite side of the BZB
by a primitive reciprocal lattice
vector $\bm G^0_{n\alpha i}$.  
The contribution to
the integral (\ref{central}) from 
conjugate BZB intersections is 
$\phi_{n\alpha i}\bm G^0_{n\alpha i}/2\pi$,
which can be apportioned equally to the ``$+$'' and
``$-$'' paths.

Eliminating the integrals over the BZB allows
band indices $n$ to be dropped, and
$\sum_n \bm K_n$  can be written (modulo $\bm G$) as a sum
$\sum_{\alpha} \bm K_{\alpha}$ of 
Fermi-surface
integrals
\begin{equation}
\bm K_{\alpha} =
\frac{1}{2\pi}
\int_{S_{\alpha}} d^2\mathcal F \, \bm k_F+
\frac{1}{4\pi}\sum_i \bm G_{\alpha i} 
\int_{\partial S_{\alpha}^i} d\mathcal A,
\label{best}
\end{equation}
where 
$d^2\mathcal F$ $\equiv$
$\mathcal F_{\mu\nu}(\bm s)ds^{\mu}\wedge ds^{\nu}$
is the Berry curvature 2-form and
$d\mathcal A$ is the connection 1-form $\mathcal A_{\mu}(\bm s) ds^{\mu}$;
$\partial S_{\alpha}^i$ are the 1-manifolds where $S_{\alpha}$
intersects the BZB, across which $\bm k_F(\bm s)$ jumps by 
$\bm G_{\alpha i}$.   These boundary terms are 
Berry-gauge dependent,
but if $S_{\alpha}$ is
nonchiral, the BZ can be chosen so all 
$\partial S^i_{\alpha}$ are closed paths, and
the gauge ambiguity is an (unknown) quantized QHE
contribution.  (If $S_{\alpha}$ is chiral, (\ref{best}) is valid
in a Berry gauge where $\mathcal A_{\mu}(\bm s)$ is periodic, and the
boundary terms cancel.) 

The integral of $1/2\pi$ times the  Berry curvature over Fermi-surface
sheet $\alpha$ is an integer Chern number $C_{\alpha}$.
If a gauge transformation shifts $\bm k_F(\bm s)$ by a constant
$\delta \bm k_F$, $\bm K_{\alpha}$ changes by
$C_{\alpha}\delta \bm k_F$.   Gauge invariance requires that the
total sum of Fermi surface Chern numbers must vanish.

If a nonchiral compact  piece
of Fermi surface has nonzero Chern number, it must enclose
a source of  Berry curvature, and 
has a hidden ``wormhole'' connection through which
``spectral flow'' can occur.  Consider
two bands linked by a ``Berry flux loop''
through a pair of Dirac points, one
below the Fermi level, the
other above  it ({both}
time-reversal and  inversion symmetries must be  broken). 
A particlelike
Fermi surface with Chern number  $C$ = $\pm 1$ surrounds
the lower Dirac point, and a holelike one with
opposite-sign  Chern number surrounds the other.
A ``spectral flow'' process driven by spatial inhomogeneities can
``pump'' states
into a band through one Dirac point and out through
the other, conserving the total number
of states per band.  States that flow
carry their occupations with them:
the volumes of the
holelike and particlelike Fermi
surfaces 
shrink or expand by the same amount, conserving
total charge.

The additional Berry phase terms in (\ref{best}) are there for a
very concrete reason:  the choice of the Brillouin zone defined
by $P_{\text{BZ}}(\bm k)$ is yet another kind of arbitrary gauge choice:
since $\bm k_F(\bm s)$ is defined to be in the BZ, however it
is chosen, there must be BZB lines on a multiply connected surface across which
$\bm k_F(\bm s)$ jumps discontinuously back into the BZ.
{The Berry phase counterterms merely guarantee that the value of
 $\bm K_{\alpha}$ is unchanged by any continuous deformation of the
standard BZ into any other primitive cell}.

The Hall conductivity also controls the
the charge density $\rho_e$ =
$e\delta \rho$ induced when a uniform magnetic
flux density is applied, keeping the Fermi energy $\mu$ fixed:
\begin{equation}
\lim_{\bm B\rightarrow 0}\left .
\frac {\partial \rho_e}{\partial F_{ab}} \right |_{\mu}
= \sigma^{ab}_0(\mu) ,  \quad F_{ab} \equiv \epsilon_{abc}B^c.
\label{streda}
\end{equation}
Such a Streda-type formula\cite{streda}
should hold separately for each Fermi surface sheet.
This can easily be found from  semiclassical quantization
with a Fermi-surface Berry phase.
For  $\bm B$ = $B\hat{\bm n}$ (technically parallel
to a lattice translation), let the set of Fermi surface
orbits in the $\bm k$-space plane $\hat {\bm n}\cdot \bm k$ 
= $k$ have cross-sectional areas $A_{\alpha}(k)$
and Berry phases $\phi_{\alpha}(k)$. 

Recall that $\sigma^{ab}_0$ is parameterized by $\bm K_{\alpha}$:
here
\begin{equation}
\hat{\bm n}\cdot \bm K_{\alpha}
= \frac{1}{2\pi}\int_{-\infty}^{\infty}  
dk\,  P_{\text{BZ}}(k\hat {\bm n}) \phi_{\alpha}(k).
\end{equation}
The semiclassical quantization condition 
is
\begin{equation}
A_{\alpha}(k_{F\alpha }^{i})
\ell_B^2 - \phi_{\alpha}(k_{F\alpha }^{i})
= 2\pi (n_{\alpha }^i + \textstyle{\frac{1}{2}}),
\end{equation}
where $\ell_B^2$ = $\hbar/|eB|$, and $n_{\alpha }^{i}$ are integers.
The $k^i_{F\alpha}$
(ordered so $k^{i}_{F\alpha}$ $> $ $k^{j}_{F\alpha}$
for $i > j$) 
are  the  Fermi momenta of  
sets of 1D Fermi gases of particles on sheet $\alpha$,
moving along field lines with a 2D density
$(2\pi\ell_B^2)^{-1}$
(one line per flux quantum).
The induced charge density is completely determined by  Fermi
surface geometry:
\begin{equation}
\rho_e = 
\frac{e}{(2\pi)^3}
\sum_{\alpha i} 
{\textstyle\frac{1}{2}}(
A_{\alpha}^{i}+ A_{\alpha}^{i+1})\Delta k_{\alpha}^{i,i+1}
 - v_{\alpha}^{i,i+1},
\end{equation}
where $\Delta k_{\alpha}^{i,i+1}$ = $k_{F\alpha}^{i+1}-k_{F\alpha}^i$,
$A^i_{\alpha}$ =
$A_{\alpha}(k_{F\alpha}^i) - (eB/\hbar)
\phi_{\alpha}(k_{F\alpha}^i)$,
and $v_{\alpha}^{i,i+1}$ is the integral of $A_{\alpha}(k)$ from
$k_{F\alpha}^i$ to $k_{F\alpha}^{i+1}$.
The Streda formula (\ref{streda}) is now easily
obtained when $B\rightarrow 0$.
Note the  differences between density and
energy shifts due to Landau quantization:
the latter derive from changes 
to states deep below the Fermi level,
and Landau diamagnetism is  {not} a Fermi surface effect.

In summary, the intrinsic
Hall conductivity of a metal with broken time-reversal
symmetry can be written as $\sigma^{ab}_0$ = $(e/2\pi \Phi_0)\epsilon^{abc}K_c$,
$\Phi_0$ = $h/e$;
I have shown that the {non-quantized} part
of the wavevector $\bm K$ ({i.e.}, the part
modulo a reciprocal lattice vector $\bm G$)
is a {topological Fermi surface property} 
given by a sum of terms $\bm K_{\alpha}$ (\ref{best}) associated
with each distinct Fermi-surface sheet.
Separate
 {adiabatically-conserved 
currents} are
associated with each such sheet
(or group of sheets mutually coupled by ``wormholes''): 
this  generalizes the ``extra'' conservation laws at
each distinct chiral Fermi point 
of a 1D Luttinger liquid.
In the absence of BCS pairing processes,
breakdown of these extra conservation laws can only occur
through non-adiabatic impurity
or surface scattering.
A ``{topological Fermi
liquid theory}'' that includes
Berry phases of  quasiparticles adiabatically
moving
on Fermi surface manifolds
must now be developed.

This work was supported in part by the U. S. National
Science Foundation (under MRSEC Grant No. DMR02-13706) at  the Princeton
Center for Complex Materials, and under Grant No. PHY99-07949 at the Kavli
Institute for Theoretical Physics, UC Santa Barbara.

\textit{Note Added:}
The simplicity of the Fermi-surface formula (\ref{best})
strongly suggests that it is a fundamental Fermi-liquid property,
also valid in an interacting system, like the relation between 
electron density and Fermi surface volume [used above
to derive (\ref{streda})].  As I will describe
elsewhere, this is indeed the 
case: the key point is that (only)
{at the Fermi surface}, the Bloch states 
$|\psi_n(\bm k_F(\bm s))\rangle$ retain their
meaning as the {zero-mode eigenfunctions} of the inverse
(exact) one-electron propagator
$G^{-1}(\bm k_F(\bm s),\omega = 0)$, which in interacting Fermi-liquid theory
remains Hermitian at $T=0$.
The formal proof uses a 3D version
of the Ward-Takahashi identity used previously
to relate the 2D integer QHE Hall conductance
to the exact (interacting)  one-electron propagator\cite{ishikawa}.
I thank P. B. Wiegmann for bringing Ref.\cite{ishikawa} to my attention.


\begin{thebibliography}{10}
\bibitem{kramers}
Kramers degeneracy with spin-orbit
coupling and inversion symmetry
leads to {non-Abelian} $SU(2)$ Berry curvature on the
Fermi surface, to be described elsewhere.
\end{thebibliography}
\end{document}